\documentclass[nofootinbib,prd,superscriptaddress,twocolumn]{revtex4}

\usepackage{amsmath}
\usepackage{amsfonts}
\usepackage{amssymb}
\usepackage{graphicx}
\usepackage[titletoc]{appendix}
\usepackage{color}
\usepackage{hyperref}
\usepackage{cleveref}
\usepackage[rightcaption]{sidecap}
\usepackage{comment}
\usepackage{soul}
\usepackage{cancel}

\usepackage{dcolumn}

\usepackage{longtable}
\usepackage{floatrow}
\usepackage{array}
\usepackage{ctable}
\usepackage{multirow}
\usepackage{siunitx}
\usepackage{tabularx}
\usepackage{booktabs}
\usepackage{supertabular}

\graphicspath{{Graphics/}}

\def\be{\begin{equation}}
\def\ee{\end{equation}}
\def\bea{\begin{eqnarray}}
\def\eea{\end{eqnarray}}

\definecolor{vividviolet}{rgb}{0.62, 0.0, 1.0}
\definecolor{amaranth}{rgb}{0.9, 0.17, 0.31}
\definecolor{palatinateblue}{rgb}{0.15, 0.23, 0.89}
\definecolor{brightpink}{rgb}{1.0, 0.0, 0.5}
\definecolor{cornflowerblue}{rgb}{0.39, 0.58, 0.93}
\definecolor{deepcarminepink}{rgb}{0.94, 0.19, 0.22}
\definecolor{radicalred}{rgb}{1.0, 0.21, 0.37}

\hypersetup{ linktoc=all,
    colorlinks, linkcolor={palatinateblue},
    citecolor={brightpink}, urlcolor={amaranth}
}

\begin{document}

\title{Parameterizing quasi-quintessence and quasi-phantom fields without the nearly flat potential approximation}

\author{Anna Chiara Alfano}
\email{a.alfano@ssmeridionale.it}
\affiliation{Scuola Superiore Meridionale, Largo S. Marcellino 10, 80138 Napoli, Italy.}
\affiliation{Istituto Nazionale di Fisica Nucleare (INFN), Sezione di Napoli Complesso Universitario Monte S. Angelo, Via Cinthia 9 Edificio G, 80138 Napoli, Italy.}

\author{Youri Carloni}
\email{youri.carloni@unicam.it}
\affiliation{Universit\`a di Camerino, Divisione di Fisica, Via Madonna delle carceri 9, 62032 Camerino, Italy.}
\affiliation{INFN, Sezione di Perugia, Perugia, 06123, Italy.}
\affiliation{INAF - Osservatorio Astronomico di Brera, Milano, Italy.}

\begin{abstract}
An alternative dark energy description based on a generalized K-essence scenario is here explored. In particular, we consider a \emph{quasi-quintessence} and/or \emph{quasi-phantom} field, whose pressure does not depend on the kinetic energy, firstly discussed in the context of the cosmological constant problem. In so doing, we fix the background evolution and investigate the main observational signatures of its corresponding fluid-like representation. The corresponding scalar field can be parameterized independently from the potential form and without imposing the condition $\omega \sim -1$ used for quintessence and phantom fields. Additionally, we constrain the model parameters by performing  Monte-Carlo Markov chain simulations through the adoption of the Metropolis-Hastings algorithm and perform separated analyses, employing different data catalogs. More precisely, as data sets we employ observational Hubble data, type Ia supernovae and the second data release from the DESI Collaboration, namely DESI DR2. We define a hierarchy among analyses and, precisely, in the first we adopt all three samples, while the second excludes the DESI data points, with the aim of facing its effect on corresponding bounds. Our findings suggest that the \emph{quasi-quintessence} scenario prefers Planck's value of the Hubble constant $H_0$, but suggesting that, when the DESI sample is excluded from our computations, $\omega_0$ enters the phantom regime, although  still compatible at $1$-$\sigma$ confidence level with a cosmological constant. Remarkably, these results appear in tension than those found for a standard quintessence, explored within the context of the recent DESI release, likely indicating that the DESI data may furnish inconclusive results depending on the kind of scalar field involved into the computation.
\end{abstract}


\maketitle
\tableofcontents

\section{Introduction}

Observational tests carried out in recent years suggest that the universe is currently undergoing a phase of accelerated expansion caused by the presence of an exotic fluid labeled dark energy \cite{1998AJ....116.1009R, 1999ApJ...517..565P,Aviles:2012ay}. This behavior seems well-described by the $\Lambda$CDM model, where $\Lambda$ denotes the dark energy fluid acting as a cosmological constant \cite{1992ARA&A..30..499C, 2001LRR.....4....1C, 2003RvMP...75..559P} and making up $\sim 70\%$ of the universe energy budget with a negative equation of state causing the universe to speed up and CDM indicates cold dark matter which accounts, together with baryonic matter, $\sim\ 30\%$ of the remaining energy budget. Although this paradigm has proven remarkably successful, it faces significant challenges, such as the cosmological constant problem\footnote{From this issue other two follow, the so-called fine-tuning and coincidence problems. The former is a discrepancy of $10^{121}$ order of magnitudes between the observed and theoretical value of $\Lambda$ energy density while the latter refers to the strange comparability between the densities of $\Lambda$ and matter at present times, i.e. $\Omega_{\Lambda_0}/\Omega_{m0}\approx 2.3$ \cite{2006IJMPD..15.1753C}.} and persistent cosmological tensions \cite{1989RvMP...61....1W, 2021APh...13102605D, 2021APh...13102604D}. As a result, various alternative models have been proposed to replace the cosmological constant with a dynamical form of dark energy \cite{2001IJMPD..10..213C, 2003PhRvL..90i1301L, 2006IJMPD..15.1753C, 2013CQGra..30u4003T, 2001PhRvD..63j3510A, 2001PhLB..511..265K, 2015EPJP..130..130C, 2025JCAP...05..034W}, spanning from extensions of Einstein's gravity \cite{2008GReGr..40..357C,2022GReGr..54...44S}, additional barotropic fluids \cite{2006IJMPD..15.1753C,2013FrPhy...8..828L,2020PhR...857....1F,2011CoTPh..56..525L,2014arXiv1401.0046M,2018RPPh...81a6901H}, unified dark energy models \cite{2025arXiv250409987L,2024PDU....4601563D,2024PhRvD.109b3510D,2018PDU....20....1C,Capozziello:2018mds,Boshkayev:2019qcx} to reconstructions and emergent phenomena \cite{Capozziello:2022jbw,Luongo:2015zgq,Luongo:2015zaa,Luongo:2010we,Luongo:2014qoa,Giambo:2020jjo}. 

In particular, recent results from the DESI Collaboration appear to favor a \emph{dynamical} dark energy scenario over a constant $\Lambda$ \cite{2025JCAP...02..021A, 2025arXiv250314738D}.

For this reason, we propose to use an alternative scalar field framework  to solve the cosmological constant problem \cite{2018PhRvD..98j3520L}, describing at the same time a matter fluid that emerges from a first-order phase transition and can speed the universe up. In this scheme, vacuum energy is canceled out as a result of the phase transition, leading to a fluid with non-zero pressure that currently drives the accelerated expansion of the universe, when it dominates over matter at late-times. In this work, we propose to use the so-called \emph{quasi-quintessence} or \emph{quasi-phantom field}, depending on the sign of the kinetic term. Then, we derive a general parameterization for this scalar field that is independent of the specific scalar potential and does not rely on $\omega\sim -1$ condition commonly assumed in the literature to parameterize quintessence and phantom fields \cite{2008PhRvD..77h3515S, 2008PhRvD..78f7303S, 2009PhRvD..79j3005D, 2009PhRvD..79h3517C, 2011PhLB..704..265D}. Afterwards, we compute a Monte-Carlo Markov chain (MCMC) analysis adopting the Metropolis-Hastings algorithm \cite{1953JChPh..21.1087M, 1970Bimka..57...97H} to derive the best-fit parameters by performing two analyses, labeled as Analysis 1 and Analysis 2, respectively. In the former we adopt a combination of observational Hubble data (OHD), the reduced Pantheon catalog of type Ia supernovae (SNe Ia) \cite{2018ApJ...853..126R} and the baryonic acoustic oscillations (BAO) from the second data release (DR2) of the DESI Collaboration \cite{2025arXiv250314738D} while in the latter we remove the DR2 data points to assess their effect on the parameters. Specifically, we find out that the exclusion of the DR2 leads to lower values of all the parameters involved in our analysis but with higher error bars. Also, the outcomes drawn out from adding the DESI-BAO is a compatibility at $2$-$\sigma$ between our value of the Hubble constant $H_0$ and the one inferred from the Planck Collaboration ($H_0=67.36\pm 0.54\ \text{km/s/Mpc}$) \cite{2020A&A...641A...6P} while no compatibility has been found with the $H_0$ from Riess ($H_0=73.04\pm 1.04$\ \text{km/s/Mpc}) \cite{2022ApJ...934L...7R}. On the other hand, this compatibility rises at $1$-$\sigma$ for Planck and at $2$-$\sigma$ for Riess when the BAO from the DESI Collaboration are excluded. Considering the energy density $\Omega_{\phi_0}$ at present times of the scalar field both our analyses indicate a compatibility at $1$-$\sigma$ with $\Omega_\Lambda=0.6847\pm 0.0073$ \cite{2020A&A...641A...6P}. Finally, the barotropic factor $\omega_0$ seems better constrained for Analysis 1, when the DR2 from the DESI Collaboration \cite{2025arXiv250314738D} are considered showing a $1$-$\sigma$ agreement with $\omega_0=-1$. When removing the DESI data points we find that $\omega_0$ seems to prefer a value entering the phantom regime, i.e. $\omega_0 <-1$ even though still compatible at $1$-$\sigma$ with $\omega_0=-1$ due to the higher error bars.

This work is divided as follows. In Sect. \ref{sez1}, we outline the theoretical framework of our model while Sect. \ref{sez2} deals with the introduction of a general parameterization of the scalar field, without imposing any \emph{a priori} assumptions. In Sect. \ref{sez3}, we constrain the fluid dynamics through two analyses, Analysis 1 and Analysis 2, performed with and without the inclusion of the DESI DR2 dataset, respectively. Finally, in Sect. \ref{sez4} we present our conclusions.

\section{Theoretical setup}\label{sez1}

The \emph{quasi-quintessence} model was first proposed in terms of barotropic fluids only, exploring the idea that matter can exhibit a non-zero pressure \cite{Luongo,Luongo:2012dv}. In this respect, the fluid responsible to characterize the universe today was barotropic and was not composed of a matter plus cosmological constant, but made of a single contribution \cite{2016PhRvD..94h3525D}. The corresponding picture yields an emergent cosmological constant \cite{Luongo:2014nld}, whose fundamental origin is not related to vacuum energy. 

With the aim of solving the cosmological constant problem, namely to reconcile the aforementioned outcomes with quantum field theory, the quasi-quintessence Lagrangian was finally introduced in Ref.~\cite{2018PhRvD..98j3520L}, where a fundamental representation of a scalar field mimicking matter was provided. 

This outcome extended the modification of gravity, presented in Ref. \cite{2010PhRvD..81d3520G}, where erroneously the authors claimed that there was no fundamental representation, through scalar field Lagrangian, for having a zero sound speed. 

The basic idea was to introduce a Lagrange multiplier, without directly invoking the need of dark energy but, again, claiming that matter with pressure may act as one single fluid characterizing the entire cosmological dynamics and \emph{de facto} conflicting with previous approaches in which a similar fluid was presented but to justify a dusty dark energy counterpart \cite{Lim:2010yk}.

The underlying fundamental theory requires a mechanism of vacuum energy cancellation, that agrees with the Weinberg no-go theorem \cite{Weinberg:1988cp,2003PhR...380..235P} as it requires a phase transition induced by a symmetry breaking. 

Consequently, the metastable phase was thus explored, namely the phase \emph{during} the aforementioned transition, in which the effects due to the potential are absolutely unavoidable \cite{2022CQGra..39s5014D}. 

Thus, during the first-order phase transition, the model triggers a cancellation mechanism that deletes the vacuum energy resolving the cosmological constant problem, but during the transition the model itself can explain inflationary dynamics. 

Accordingly, the underlying scalar field represents matter that, as consequence of the transition, exhibits a net and not fine-tuned constant term, interpreted as source to speed the universe up at late-times, but at primordial epoch, during the transition, provides an inflationary period, unifying late and early stages of universe evolution into a single scheme. 

The model was thus generalized at all stages, invoking the fact that, in principle, as well as standard quintessence, the choice of the underlying potential may furnish a different fluid \cite{Luongo:2023jnb}. 

Moreover, the model was explored in view of inflation \cite{Luongo:2024opv}, particle production \cite{Belfiglio:2023rxb,2023CQGra..40j5004B} and structure formation \cite{2011PhRvD..84h9905A}, showing very promising results. 

The basic demands is that the net single fluid shows an equation of state that appears the same of the \emph{total} equation of state computed by two fluids, at least, in the $\Lambda$CDM paradigm. In this respect, the degeneracy between the two paradigms was evident, suggesting that the quasi-quintessence constituent acts as a dark fluid, with constant pressure and evolving equation of state and density \cite{Luongo:2014nld,Luongo:2011yk,Aviles:2014mua}.

In this work, in view of the DESI results, it is natural to resort the quasi-quintessence fluid, to explore the chance of breaking the degeneracy with the $\Lambda$CDM scenario.

Hence, we revisit the aforementioned approach in light of recent DESI results \cite{2025JCAP...02..021A, 2025arXiv250314738D}, proposing the same Lagrangian to describe a dynamical dark energy, namely 
\begin{equation}
    \mathcal{L}=K(X,\phi)+\lambda Y[X,\nu(\phi)]-V(X,\phi),
\end{equation}
in which $X$, $Y$ are kinetic functions and $\lambda$ is a Lagrange multipler. Here, the Lagrangian depends on the scalar field $\phi$ and its kinetic term $X=g^{ab}\nabla_a \phi \nabla_b \phi/2$, with $g^{ab}$ denoting the metric tensor, and $\nu(\phi)$ an inertial mass. At this stage, by varying the action with respect to $\lambda$, $\phi$ and $g^{ab}$, we obtain
\begin{subequations}
    \begin{align}
    &Y=0,\\
    &\mathcal{L}_{,\phi}-\nabla_{a}(\mathcal{L}_{,X}\nabla^{a}\phi)=0,\\
    &T_{ab}=\mathcal{L}_{,X}\nabla_{a}\phi\nabla_{b}\phi-(K-V)g_{ab},\label{eq:EM}
\end{align}
\end{subequations}
with $\mathcal{L}_{,X}=K_{,X}-V_{,X}+\lambda Y_{,X}$ and $\mathcal{L}_{,\phi}=K_{,\phi}-V_{,\phi}+\lambda Y_{,\nu}\nu_{\phi}$, where the subscripts with a preceding comma denote partial derivatives. Then, by adopting the definition of the four-velocity as $v_{a}=\nabla_{a}\phi/\sqrt{2X}$, the energy-momentum tensor introduced in Eq. \eqref{eq:EM} can be reformulated as 
\begin{equation}
    T_{ab}=2X\mathcal{L}_{,X}v_{a}v_{b}-(K-V)g_{ab},\label{eq:EM2}
\end{equation}
from which the energy density and the pressure are obtained as
\begin{align}
    &\rho_{\phi}=2X\mathcal{L}_{,X}-(K-V),\\
    &P_{\phi}=K-V,
\end{align}
respectively. As shown in Refs.~\cite{2018PhRvD..98j3520L, 2022CQGra..39s5014D}, the pressure remains constant before and after the first-order phase transition, implying that the kinetic function assumes a constant value, i.e., $K=K_0=\text{const}$, so the energy density and the pressure becomes respectively  $\rho_{\phi}=2X\mathcal{L}_{,X}+\mathcal{V}$ and $P_{\phi}=-\mathcal{V}$, where $\mathcal{V}=V-K_{0}$ represents the new potential\footnote{From now on,  we identify $V$ with $\mathcal{V}$, as they differ only by an additive constant and can thus be used interchangeably for our purposes.}.

This formalism can now be applied to model an alternative dark energy field by assuming $2X\mathcal{L}_{,X} = \varepsilon \dot{\phi}^2 / 2$, where $\varepsilon = +1$ indicates the \emph{quasi-quintessence}, while $\varepsilon = -1$ characterizes the \emph{quasi-phantom field}. 

It is worth noting that this approach may be favored over conventional dynamical dark energy models, such as quintessence or phantom fields, particularly from the perspective of structure formation. Indeed, while standard scalar field prescriptions predict $c_s = 1$ as sound speed, our formalism leads to a vanishing sound speed $c_s = 0$, making this model degenerate with the $\Lambda$CDM paradigm.

\section{Model parameterization}\label{sez2}

The evolution of the field $\phi$ in the case of \emph{quasi-quintessence} is given by the Klein-Gordon equation \cite{2010PhRvD..81d3520G, 2022CQGra..39s5014D}
\begin{equation}\label{kgeq}
    \ddot{\phi}+\frac{3}{2}H\dot{\phi}+\varepsilon{V}^\prime(\phi)=0,
\end{equation}
where the prime represents the derivative with respect to $\phi$ and $\varepsilon$ varies according to if we are considering the \emph{quasi-quintessence} ($\varepsilon=+1$) or \emph{quasi-phantom} ($\varepsilon=-1$) field.

In this alternative representation of the scalar field, the kinetic term does not appear in the pressure, i.e. $P_\phi=-V$ due to the cancellation mechanism of the cosmological constant during a first-order phase transition while the energy density is $\rho_\phi=\frac{1}{2}\varepsilon\dot{\phi}^2+V$.

Now, if we consider the derivative of the barotropic factor $\omega^\prime=d\omega/d\ln(a)=H^{-1}d\omega/dt$, where $a$ is the scale factor, Eq. \eqref{kgeq} can be re-written as \cite{1999PhRvD..59l3504S, 2006PhRvD..73f3501C, 2008PhRvD..77h3515S, 2008PhRvD..78f7303S}
\begin{equation}\label{kgpot}
    \mp\frac{{V}^\prime}{{V}}=\left(1+\frac{1}{3}\frac{d\ln x}{d\ln a}\right)\sqrt{\frac{3}{2}\frac{(1+\omega)\varepsilon}{\Omega_\phi}},
\end{equation}
where we used $H=\sqrt{\rho_\phi/3\Omega_\phi}$ and $x=-(1+\omega)/\omega$. 

Following the prescription in Ref. \cite{2008PhRvD..77h3515S} we introduce an additional parameter $\beta=\varepsilon(1+\omega)$ and set $\lambda=-{V}^\prime/{V}$. In this way, we can simplify Eq. \eqref{kgpot} as
\begin{equation}\label{betaprime}
    \beta^\prime=\left(\varepsilon\beta-1\right)\lambda\sqrt{6\beta\Omega_\phi}-3\beta\left(\varepsilon\beta-1\right),
\end{equation}
Afterwards, we consider the evolution of the fractional density of dark energy 
\begin{equation}\label{evde}
    \Omega^\prime_\phi=3\left(1-\varepsilon\beta\right)\Omega_\phi(1-\Omega_\phi).
\end{equation}
Taking the ratio between Eqs. \eqref{betaprime}-\eqref{evde} we end up with
\begin{equation}\label{der2}
    \frac{d\beta}{d\Omega_\phi}=\frac{\beta}{\Omega_\phi(1-\Omega_\phi)}-\sqrt{\frac{2}{3}}\frac{\lambda_0}{(1-\Omega_\phi)}\sqrt{\frac{\beta}{\Omega_\phi}},
\end{equation}
where $\lambda=\lambda_0=-V^\prime/V\Big|_{\phi=\phi_0}$ as the value of $\lambda$ at the initial value of the scalar field $\phi_0$ before it begins to roll down the potential. This follows from the slow-roll conditions
\begin{equation}
    \left(\frac{V^\prime}{V}\right)^2\ll 1,\quad \frac{1}{V}\frac{d^2V}{d\phi^2}\ll 1.
\end{equation}
Now, focusing our attention on Eq. \eqref{der2} it appears a more general expression without invoking the approximation $\beta\ll 1$ typically employed in the standard scalar field framework \cite{2008PhRvD..77h3515S, 2008PhRvD..78f7303S}. This indicates that the alternative scalar field formulation enables a broader parameterization of the scalar field dynamics, rather than being restricted to the specific case where a nearly flat potential is assumed. Moreover, it is worth to stress that Eq. \eqref{der2} does not depend on $\varepsilon$ making the expression valid for both the \emph{quasi-quintessence} and \emph{quasi-phantom} field. This last result was already found in Refs. \cite{2008PhRvD..77h3515S, 2008PhRvD..78f7303S} for the quintessence scenario but with the imposition of the \emph{nearly flat potential approximation}.

At this point, we solve Eq. \eqref{der2} by setting $\beta=s^2$ obtaining
\begin{equation}
    2\frac{ds}{d\Omega_\phi}=\frac{s}{\Omega_\phi(1-\Omega_\phi)}-\sqrt{\frac{2}{3}}\frac{\lambda_0}{\sqrt{\Omega_\phi}(1-\Omega_\phi)}.
\end{equation}
Considering as boundary conditions $s(\Omega_\phi=1/2)=0$ and then restoring $s^2=\beta=\varepsilon(1+\omega)$ we end up with
\begin{align}\label{w1}\nonumber
    &\varepsilon(1+\omega)=\frac{2\Omega_\phi\lambda_0}{3(1-\Omega_\phi)}\times\\&\times\left[\tanh^{-1}(\sqrt{1-\Omega_\phi})-\tanh^{-1}\left(\frac{1}{\sqrt{2}}\right)\right]^2.
\end{align}
Now, our goal is to write the previous equation in terms of the barotropic factor at present times, i.e. $\omega\equiv\omega_0$. To do that we first re-write Eq. \eqref{evde} in the limit $\omega\rightarrow-1$ 
\begin{equation}\label{evde0}
    \Omega_\phi=[1+(\Omega_{\phi_0}^{-1}-1)a^{-3}]^{-1}.
\end{equation}
Inserting Eq. \eqref{evde0} inside Eq. \eqref{w1} and normalizing $\omega$ to $\omega=\omega_0$ at present times give us
\begin{align}\label{wQQ}
\nonumber
1+\omega
&= \frac{(1+\omega_0)}{a^{-3}}
\Biggl[\tanh^{-1}\left(\sqrt{\frac{(\Omega_{\phi_0}-1)a^{-3}}{1 + (\Omega_{\phi_0}-1)a^{-3}}}\right)+
\\&\quad-\tanh^{-1}\left(\frac{1}{\sqrt{2}}\right)\Biggl]^2\times\\&\quad\Biggl[\tanh^{-1}\left(\sqrt{1-\Omega_{\phi_0}}\right)-\tanh^{-1}\left(\frac{1}{\sqrt{2}}\right)\Biggl]^{-2}.\nonumber
\end{align}

\section{Cosmic probes}\label{sez3}

For our analyses we employ three different probes as listed below.

\begin{enumerate}

    \item [-] {\bf OHD.} We adopt the most updated OHD catalog composed of 34 measurements of the Hubble rate $H(z)$ in the redshift range $z\in[0.07, 1.965]$. These data points, although they present high error bars, are derived in a totally model-independent way by adopting the so-called {\it cosmic chronometers} \cite{2002ApJ...573...37J}. These chronometers are passively-evolving red galaxies and $H(z)$ is inferred through the \emph{differential age method} considering the age difference of pairs of nearby galaxies that formed at the same time at different redshift $z$ through the relation $H(z)= -(1+z)^{-1}\Delta t/\Delta z$. For the complete sample see e.g. Refs. \cite{2024JCAP...12..055A, 2025PhRvD.111b3512C,Luongo:2024fww}.
    
    \item [-] {\bf SNe Ia.} We consider the reduced SNe Ia sample \cite{2018ApJ...853..126R} consisting in 6 correlated normalized Hubble rate data points $E(z)=H(z)/H_0$ in the redshift range $z\in[0.07, 1.5]$ and obtained within the assumption of a flat universe, i.e. $\Omega_k=0$. This sample is complitely equivalent to the Pantheon catalog \cite{2018ApJ...859..101S} with the advantage of decreasing possible computational complexities and speed up the analysis.
    
    \item [-] {\bf DESI DR2.} We use the second release of the DESI-BAO data points from the DESI Collaboration \cite{2025arXiv250314738D} consisting in measurements of the transverse comoving distance $d_M/r_d$, the Hubble rate distance $d_H/r_d$ and the angle-average distance $d_V/r_d$ given by the combination of the previous two distances within the redshift interval $z\in [0.295, 2.33]$. We adopt the $13$ data points, displayed in Tab. IV of Ref. \cite{2025arXiv250314738D}, for which the covariance matrix is known\footnote{The full covariance matrix is available in the GitHub repository of the code \emph{Cobaya}~\url{https://github.com/CobayaSampler/bao_data/blob/master/desi_bao_dr2/desi_gaussian_bao_ALL_GCcomb_cov.txt}}. For all distances the sound horizon at the baryon drag epoch is fixed to the Planck Collaboration \cite{2020A&A...641A...6P} value $r_d=(147.09\pm0.26)\ \text{Mpc}$ to break the $r_d-H_0$ degeneracy\footnote{Alternatevely, one can perform an analysis in which $r_d$ is varied within an interval in which both Planck and DESI expectations fall. Afterwards, the value of $r_d$ preferred by model selection criteria is taken \cite{2024JCAP...12..055A, 2025PhRvD.111b3512C}.}.
\end{enumerate}

The best-fit parameters for our model are found by maximizing the log-likelihood $\ln\mathcal{L}\propto-\chi^2/2$ where the $\chi^2$ varies depending on if the considered measurements are correlated or not. In particular, for our chosen probes the $\chi^2$ for each of them are given by
\begin{subequations}
\begin{align}
    \chi^2_O &= \sum_{i}^{N_O} \left[ \frac{H_i - H(z_i)}{\sigma_{H_i}} \right]^2, \\
    \chi^2_S &= \sum_{i}^{N_S} \Delta\xi_i^\text{T} \mathcal{C_S}^{-1} \Delta\xi_i, \\
    \chi^2_D &= \sum_{i}^{N_D}\Delta X_i^\text{T}\mathcal{C_D}^{-1}\Delta X_i,
\end{align}
\end{subequations}
where the subscript $O$ refers to the OHD sample with $H_i$ being the Hubble rate measurement with attached errors $\sigma_{H_i}$. On the other hand, the subscript $S$ refers to the SNe Ia catalog where $\Delta\xi_i=E_i-E(z_i)$ and $\mathcal{C_S}^{-1}$ the inverse covariance matrix. Finally, the subscript $D$ defines the DESI-BAO DR2 catalog where $\Delta X_i=X_i-X(z_i)$ with $X_i=\{d_M/r_d,\ d_H/r_d,\ d_V/r_d\}$ being the DR2 data with attached errors $\sigma_{X_i}$ and $\mathcal{C_D}^{-1}$ the inverse matrix for the DR2. For all probes $H(z_i)$, $E(z_i)$ and $X(z_i)$ are the theoretical Hubble rate, normalized Hubble rate and DESI-BAO distances.

In using these probes, we define the Hubble rate as 
\begin{equation}
    H(z)=H_0\sqrt{\Omega_{m_0}(1+z)^3+\Omega_{\phi_0} f(z)},
\end{equation}
where $\Omega_{m_0}=1-\Omega_{\phi_0}$ with the $0$ representing the aforementioned quantities at present times while
\begin{equation}
    f(z)=\exp\left[3\int^z_0\frac{1+\omega(z^\prime)}{1+z^\prime} \ dz^\prime\right],
\end{equation}
with $\omega(z^\prime)$ defined in Eq. \eqref{wQQ} where we consider the relation between the scale factor and the redshift, i.e. $a=(1+z)^{-1}$.

\subsection{Numerical outcomes}

\begin{figure}[t]
    \centering
    \includegraphics[width=1.\linewidth]{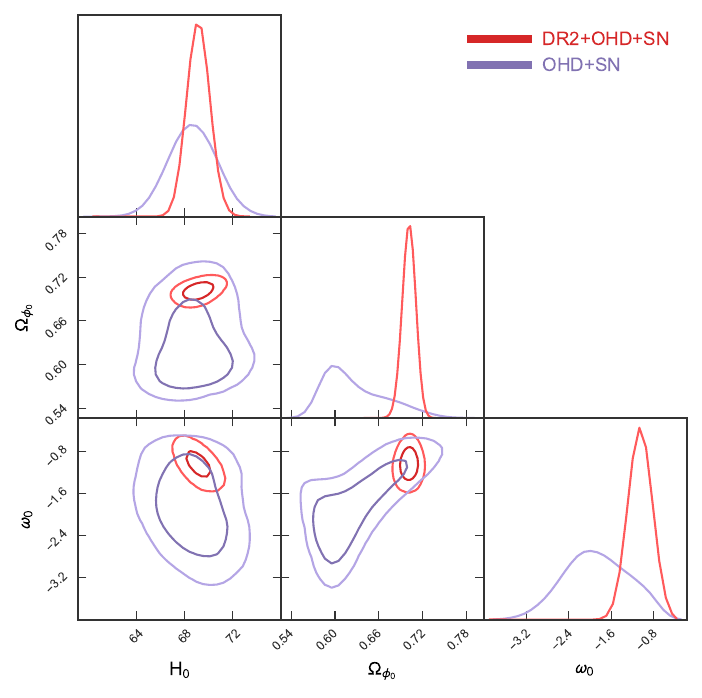}
    \caption{Contour plots of the best-fit cosmological parameters. In red we show the contour for Analysis 1 while in purple the contour for Analysis 2.}
    \label{fig:contourQQ}
\end{figure}

To derive the best-fit parameters showed in Tab. \ref{tab:bfQQ} with the contour plots depicted in Fig. \ref{fig:contourQQ} using the free Python-based code pygtc \cite{Bocquet2016} we run two MCMC analyses adopting the already discussed probes. Specifically, our analysis is divided as follows.

\begin{table}[t]
    \centering
    \setlength{\tabcolsep}{.5em}
    \renewcommand{\arraystretch}{1.}
    \begin{tabular}{ccc}
    \hline
    $H_0$ & $\Omega_{\phi_0}$ & $\omega_0$  \\
    \hline
    \multicolumn{3}{c}{Analysis 1}\\
    \hline
     $69.23^{+1.188(2.015)}_{-1.312(2.084)}$ & $0.702^{+0.012(0.019)}_{-0.014(0.023)}$ & $-1.086^{+0.339(0.520)}_{-0.238(0.409)}$\\
     \hline
     \multicolumn{3}{c}{Analysis 2}\\
     \hline
     $68.60^{+3.370(5.110)}_{-2.981(4.731)}$ & $0.597^{+0.098(0.144)}_{-0.028(0.043)}$ & $-2.027^{+1.182(1.488)}_{-0.767(1.294)}$\\
     \hline
    \end{tabular}
    \caption{Best-fit parameters with attached $1$-$\sigma$($2$-$\sigma$) errors. Upper panel shows the results for Analysis 1 while lower panel shows the results for Analysis 2 without DESI DR2 \cite{2025arXiv250314738D}.}
    \label{tab:bfQQ}
\end{table}

\begin{enumerate}
    \item [-] {\bf Analysis 1.} We consider a combination of all three data sets, i.e. OHD+SNe Ia+DESI DR2. In this case the best-fit parameters are probed by maximizing the following log-likelihood
    \begin{equation}
        \ln\mathcal{L}_1=\ln\mathcal{L}_O+\ln\mathcal{L}_S+\ln\mathcal{L}_D.
    \end{equation}
    The results inferred from Analysis 1 are displayed in the top panel of Tab. \ref{tab:bfQQ}.

    From this first analysis, we find that our $H_0$ only agrees at $2$-$\sigma$ with the expected value from the Planck Collaboration \cite{2020A&A...641A...6P}, i.e. $H_0=(67.36\pm 0.54)\ \text{km/s/Mpc}$ while no compatibility has been found when the $H_0=(73.04\pm 1.04)\ \text{km/s/Mpc}$ from SNe Ia is accounted for \cite{2022ApJ...934L...7R}.

    Regarding $\Omega_{\phi_0}$, our analysis suggests that our value is in agreement at $1$-$\sigma$ with the energy densiy inferred from the Planck Collaboration \cite{2020A&A...641A...6P} in the case of a pure cosmological constant, i.e. $\Omega_\Lambda=0.6847\pm0.0073$.

    Finally, the $\omega_0$ resulting from the MCMC computation shows a preference at $1$-$\sigma$ for $\omega_0=-1$.

    \item [-] {\bf Analysis 2.} Same as Analysis 1 but in this case we remove the DESI-BAO DR2 employing only OHD+SNe Ia to check how they affect the overall analysis. In this case the best-fit parameters are derived by maximizing the following log-likelihood
    \begin{equation}
        \ln\mathcal{L}_2=\ln\mathcal{L}_O+\ln\mathcal{L}_S.
    \end{equation}
    The results inferred from Analysis 2 are displayed in the lower panel of Tab. \ref{tab:bfQQ}.

    As done for Analysis 1 we first focus ourselves on the agreement between $H_0$ and the values derived from both Planck and SNe Ia. When the DESI-BAO are excluded from our analysis the agreement between the Planck satellite \cite{2020A&A...641A...6P} and our $H_0$ is at $1$-$\sigma$ while with Riess \cite{2022ApJ...934L...7R} is at $2$-$\sigma$.

    Focusing on the scalar field energy density, also in this case the compatibility between our $\Omega_{\phi_0}$ and Planck's value in the $\Lambda$CDM scenario is at $1$-$\sigma$.

    Also for this analysis our $\omega_0$ agrees at $1$-$\sigma$ with $-1$.
\end{enumerate}

\section{Final outlooks and perspectives}\label{sez4}

Since its first data release back in 2024 the DESI Collaboration \cite{2025JCAP...02..021A, 2025arXiv250314738D} showed a preference towards a dynamically evolving dark energy instead of a pure cosmological constant as the concordance paradigm suggests. This preference was also confirmed with the second data release \cite{2025arXiv250314738D} and since than many works have been published reinforcing or dismissing this claim \cite{2025PhRvD.111b3512C, 2024JCAP...12..055A, 2025arXiv250519029C, 2025PhRvD.112b3508G, 2025arXiv250615091K, 2024JCAP...10..035G, 2025arXiv250612004H}.

However, the model favored by DESI, the Chevallier-Polarski-Linder (CPL) parameterization \cite{2001IJMPD..10..213C, 2003PhRvL..90i1301L}, does not address the cosmological constant problem. In light of this, we propose a scalar field framework labeled as \emph{quasi-quintessence} that solves the cosmological constant problem through a cancellation mechanism leading to a fluid that speeds up the universe at late times \cite{2018PhRvD..98j3520L, 2022CQGra..39s5014D}. Particularly, we adopt the same Lagrangian as in Refs. \cite{2018PhRvD..98j3520L, 2022CQGra..39s5014D} but instead of considering the field as a matter field we reinterpret it as a dark energy scalar field.

Afterwards, we parameterize this scalar field through a barotropic factor $\omega$ that encapsules both the \emph{quasi-quintessence} and \emph{quasi-phantom} scenarios via the parameter $\varepsilon$. Our parameterization also has the advantage that unlike other ones proposed in the literature, see e.g. Refs. \cite{2008PhRvD..77h3515S, 2008PhRvD..78f7303S, 2009PhRvD..79j3005D, 2009PhRvD..79h3517C, 2011PhLB..704..265D} it does not rely on the \emph{nearly flat potential approximation} since it naturally vanishes during the calculations.

Finally, we use our parameterization to achieve constraints on the cosmological parameters. Specifically, we perform two MCMC analyses adopting the Metropolis-Hastings algorithm \cite{1953JChPh..21.1087M, 1970Bimka..57...97H} labeled as Analysis 1 and Analysis 2, respectively. For Analysis 1 we use a combination of OHD, the reduced Pantheon catalog of SNe Ia and the DESI-BAO DR2 while in Analysis 2 we exclude DESI data points to test their effect on the analysis.

The outcomes from our computations suggest that when the DESI-DR2 is accounted for we find better constraints for all the parameters involved as can be seen from the contours in Fig. \ref{fig:contourQQ}. When addressing the Hubble tension we found a major compatibility with $H_0=(67.36\pm 0.54)\ \text{km/s/Mpc}$ from the Planck Collaboration \cite{2020A&A...641A...6P}. Specifically, for Analysis 1 the compatibility is at $2$-$\sigma$ while for Analysis 2 it boosts at $1$-$\sigma$.

Regarding the other two cosmological parameters involved we find for both analyses an agreement at $1$-$\sigma$ between our $\Omega_{\phi_0}$ and $\Omega_\Lambda= 0.6847\pm 0.0073$ \cite{2020A&A...641A...6P} while $\omega_0$ agrees at $1$-$\sigma$ with $-1$ although when the DESI-BAO data points are eliminated from our analysis our $\omega_0$ seems to prefer a value towards the phantom regime $\omega_0<-1$ but still compatible at $1$-$\sigma$ with $-1$ due to the larger attached errors.

Future works will focus on investigating the behaviour of quasi-quintessence at earlier epochs by adopting high-$z$ probes such as gamma-ray bursts and standard sirens.

\section*{Acknowledgements}

ACA acknowledges the support of the Istituto Nazionale di Fisica Nucleare (INFN) Sez. di Napoli, Iniziativa Specifica QGSKY. YC acknowledges financial support from the Brera National Institute of Astrophysics (INAF). The authors are grateful to Orlando Luongo for interesting discussions on the model of quasi-quintessence and for the support throughout the preparation of this work. This paper is based upon work from COST Action CA21136 Addressing observational tensions in cosmology with systematics and fundamental physics (CosmoVerse) supported by COST (European Cooperation in Science and Technology).

\bibliography{bibliography}

\end{document}